%% file: CADENCE.tex
\documentclass[acmsmall,screen,nonacm]{acmart}
\usepackage{graphicx}
\usepackage{tcolorbox}
\usepackage{algorithm}
\usepackage{algpseudocode}
\usepackage{wrapfig}
\usepackage{enumitem}
\usepackage{multirow}
\usepackage{booktabs}
\usepackage{colortbl}
\usepackage{makecell}

\AtBeginDocument{%
  }

\begin{document}

\title{Cadence: Strategic Guidance for Coding Agents}


\author{Minxing Wang}
\affiliation{%
  \institution{Singapore Management University}
  \country{Singapore}}
\email{mx.wang.2026@phdcs.smu.edu.sg}

\author{He Ye}
\affiliation{%
  \institution{University College London}
  \country{United Kingdom}}
\email{he.ye@ucl.ac.uk}

\author{Earl T. Barr}
\affiliation{%
  \institution{University College London}
  \country{United Kingdom}}
\email{e.barr@ucl.ac.uk}

\author{Yintong Huo}
\affiliation{%
  \institution{Singapore Management University}
  \country{Singapore}}
\email{ythuo@smu.edu.sg}

\begin{abstract}
\input{Chapters/abstract}
\end{abstract}



\keywords{Coding Agents, Runtime Monitoring, Large Language Models, Self-Adaptive Scheduling}


\maketitle

\input{Chapters/Chapter1-Introduction}
\input{Chapters/Chapter2-Motivating_Study}
\input{Chapters/Chapter3-Methodology}

\input{Chapters/Chapter4-Experimental_Setup}
\input{Chapters/Chapter5-Evaluation}
\input{Chapters/Chapter6-Threats_to_Validity}
\input{Chapters/Chapter7-RelatedWork}
\input{Chapters/Chapter8-Conclusion}

\bibliographystyle{ACM-Reference-Format}
\bibliography{references}

\end{document}

%% file: Chapters/abstract.tex
Runtime monitors are increasingly used to improve the reliability of LLM-based coding agents by inspecting execution trajectories and delivering corrective guidance upon detecting misbehavior. However, their effectiveness remains limited by static guidance triggering schemes. 
Existing monitors rely either on fixed inspection intervals, missing timely guidance during severe misbehaviors while incurring unnecessary overhead during healthy execution, or on rigid heuristic rules, failing to detect complex reasoning errors.
To address these limitations, we propose Cadence, a dynamic monitoring framework that adaptively schedules inspections and delivers guidance according to the agent's real-time execution health. Cadence consists of two core modules: a two-tier intervention module and an inspection scheduler. The intervention module delivers advisory-level guidance for normal executions and minor lapses, while providing replacement-level guidance for severe misbehaviors. Driven by the intervention level, the scheduler adjusts the inspection frequency by tightening supervision after replacement-level guidance and relaxing it after advisory-level guidance. Evaluated on 300 SWE-bench Lite tasks across two distinct agents, mini-swe-agent and Moatless, Cadence achieves the highest resolve rate among all evaluated monitors. Specifically, Cadence outperforms vanilla agents by 25.33\% (+76 resolved tasks) on mini-swe-agent and 15.67\% (+47 resolved tasks) on Moatless, while maintaining competitive token efficiency compared to state-of-the-art baselines.

%% file: Chapters/Chapter1-Introduction.tex
\section{Introduction}
\label{sec:introduction}
LLM-based coding agents have been widely applied to automate software maintenance tasks, such as bug fixing and new feature implementation~\cite{swe-agent, moatless, openhands, autocoderover, agentless}. Despite their promise, agents still fail in complex coding tasks, especially long-horizon ones~\cite{multiagentfail, trail, whichandwhen, Process-oriented-error-analysis, famas}.

To improve coding agents, existing work examines step-by-step agent trajectories and provides guidance to prevent error actions.
This work can be divided into two categories based on the stage of guidance: \textit{offline learning} and \textit{online monitoring}. Offline learning relies on historical execution records to post-train the agent's backbone model~\cite{swe-rl, swe-gym, swe-smith, agentracer}, or to construct a memory bank as a domain knowledge base~\cite{swe-exp, expel, sage}. 
However, these offline methods cannot intervene during test-time execution once an agent deviates.
In contrast, runtime monitors inspect the agent's intermediate execution trace on-the-fly and inject guidance to correct misbehavior~\cite{liveplan, wink}. 
Recent studies show that early mistakes propagate into cascading failures, making dynamic intervention essential to catch errors before they become unrecoverable~\cite{multiagentfail, snowball, resilience-of-agent}.

Runtime monitoring approaches deploy a dedicated monitor to supervise the execution trajectory of a coding agent. By inspecting ongoing execution steps, the monitor
identifies misbehaviors (i.e., error steps), and injects corrective guidance into the coding agent's context. A central challenge is deciding \textit{when the monitor should deliver this guidance}. Existing approaches fall into two categories:
1) \textit{periodic guidance}, which inspects progress at fixed step intervals, then prompts corrective instructions into the context~\cite{wink}; and 
2) \textit{rule-based guidance}, which triggers on predefined heuristic rules to catch specific error patterns. When patterns appear, the monitor intervenes with operational guidance tailored to the matching failure~\cite{liveplan}.

Our preliminary study shows that neither of these static trigger mechanisms delivers guidance properly. Rule-based guidance overlooks reasoning errors, which manifest in natural language and follow no fixed misbehavior patterns. Periodic guidance, in contrast, oscillates between \textit{over-guidance} and \textit{under-guidance}. When the coding agent only encounters minor lapses, frequent interventions add unnecessary overhead (over-guidance). Conversely, when the coding agent enters severe failure actions (e.g., repeating an invalid patch), sparse guidance fails to correct the misbehavior before it compounds irreversibly (under-guidance).

To address this, we argue that monitors must move beyond rigid schedules and simple rules and instead provide \textbf{strategic guidance}: an LLM-based monitoring policy that decides both when to inspect and how to intervene based on the agent's runtime execution health. Rather than treating inspection and intervention as separate decisions, strategic guidance unifies them, dynamically tuning the inspection frequency while delivering context-aware interventions tailored to the agent's evolving health status.

Guided by this principle, we propose Cadence, a holistic runtime monitoring framework for coding agents that synergizes adaptive inspection scheduling with a two-tier intervention mechanism.
The first module is a \textbf{two-tier intervention module}, which evaluates agent's recent execution health and applies an intervention at the appropriate level: \textit{advisory} or \textit{replacement}. When the coding agent stays on track or exhibits minor lapses, an advisory-level intervention provides lightweight guidance while allowing the current step to proceed. When severe misbehaviors occur, a replacement-level intervention discards the erroneous step and prompts the agent to regenerate an alternative action.
The second module is an \textbf{inspection scheduler}, which dynamically adjusts the inspection frequency based on the first module's output. Using the resulting intervention level as a direct signal of execution health, it extends the inspection interval upon an advisory intervention to reduce monitoring overhead, but resets the interval to a base value upon a replacement intervention to enforce close oversight.
Together, these modules allow Cadence to adaptively balance intervention effectiveness while limiting unnecessary monitoring burden.

We instantiate Cadence for autonomous bug-fixing agents, though its strategic guidance mechanism generalizes to broader coding tasks because it does not rely on task-specific heuristics.
We comprehensively evaluate Cadence on SWE-bench Lite~\cite{swebench} across different scaffolds and backbones. Our results show that Cadence consistently outperforms existing runtime monitors, improving the task resolve rate by 25.33\% (+76 resolved tasks) over the unmonitored baselines in mini-swe-agent, and 15.67\% (+47 tasks) in Moatless. Furthermore, Cadence demonstrates competitive token efficiency against existing monitors, reducing average total tokens per task to 0.48M on Moatless and 1.03M on mini-swe-agent, compared with 0.51M and 1.47M for state-of-the-art monitor.

To sum up, the main contributions of this paper are threefold:
\begin{itemize}
    \item We conduct a preliminary study on existing runtime monitors and find that rule-based guidance misses reasoning errors that follow no fixed pattern, while periodic guidance oscillates between over-guidance and under-guidance.
    \item To the best of our knowledge, we propose Cadence, the first runtime monitor that strategically inspects and guides coding agents, by combining a two-tier intervention module and an inspection scheduler to adjust inspection frequency according to the agent's execution health.
    \item We comprehensively evaluate Cadence on SWE-bench Lite, showing that Cadence improves resolve rates by up to 25.33\% over vanilla coding agents while achieving up to 29.93\% token savings compared with state-of-the-art monitors.
\end{itemize}

%% file: Chapters/Chapter2-Motivating_Study.tex
\begin{figure}[t]
\centering
\includegraphics[width=0.9\linewidth]{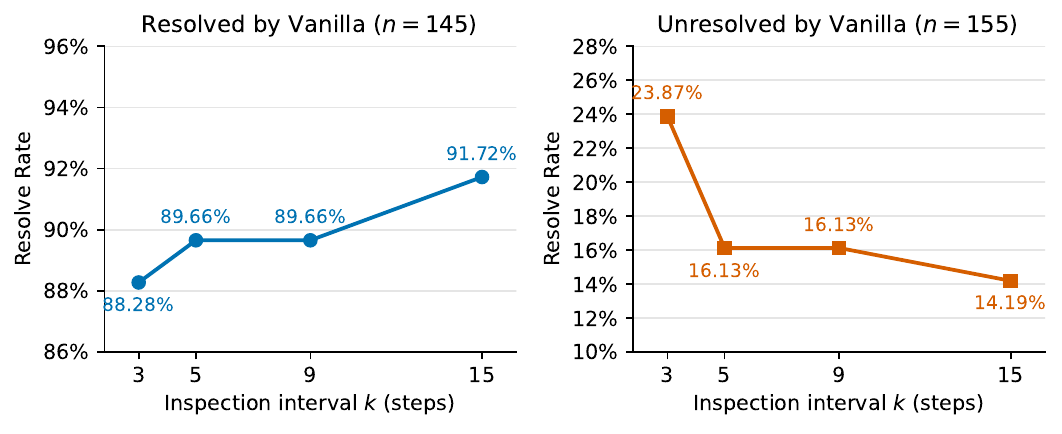}
\caption{Task resolve rate across different inspection intervals $k$ for tasks resolved ($n=145$) and unresolved ($n=155$) by the vanilla agent. Note that the two subplots use different $y$-axis scales.}
\label{fig:preliminary-study}
\end{figure}

\section{Motivating Study}

Existing runtime monitors deliver guidance through static triggers, relying on either fixed inspection intervals or predefined heuristic rules. In this section, we empirically examine the limitations of both mechanisms on SWE-bench Lite. 

\subsection{Periodic Guidance: Does One Interval Fit All?}
\label{sec:motivating study}
Existing monitors with periodic guidance focus on refining prompts to detect agent misbehaviors, while simply adopting a fixed inspection interval. To investigate the impact of such periodic guidance on agent performance, we implement a basic fixed-interval monitor over the mini-swe-agent framework. In this study, the underlying coding agent employs GPT-5-mini as its backbone, whereas the monitor utilizes GPT-5.2. At each inspection point, the monitor prompts an LLM to inspect the problem description and the agent's recent execution trajectory, generating targeted guidance that is appended to the agent's context for subsequent execution.

To examine how inspection frequency interacts with agent runtime behavior, we partition SWE-bench Lite into two distinct groups based on the outcome of the vanilla (i.e., without monitoring) mini-swe-agent: 145 resolved tasks and 155 unresolved tasks. Manual inspection of all 300 vanilla trajectories by two experts reveals that two groups represent distinct execution health profiles.
The resolved group represents scenarios where the agent stays on track or recovers from minor lapses autonomously. In contrast, the unresolved group reflects compromised execution health, where the agent suffers from severe misbehaviors that compound into irrecoverable failure. 

We evaluate four distinct inspection intervals, namely 3, 5, 9, and 15 steps, plotting the resolve rate against the interval for each group in Figure~\ref{fig:preliminary-study}. As the inspection interval increases, the resolve rate exhibits \textit{opposing trends} between the two groups. For the group with compromised execution health, a larger interval degrades performance, dropping the resolve rate from 23.87\% at interval 3 to 14.19\% at interval 15, representing a severe 9.68\% decline. 
In contrast, for the group with favorable execution health, the resolve rate consistently remains high across all interval settings, staying at or above 88.28\%.
Meanwhile, increasing the interval from 3 to 15 monotonically reduces token consumption in both groups, moving from 0.45M to 0.30M tokens for the resolved group and from 0.70M to 0.49M tokens for the unresolved group. These findings demonstrate that no single fixed interval fits all executions. Healthy executions benefit from sparser checks that minimize disruption and token overhead, whereas compromised executions require tight supervision to prevent failure propagation.

\begin{tcolorbox}[colback=blue!5, colframe=blue!50!black, boxrule=0.5pt, left=6pt, right=6pt, top=4pt, bottom=4pt]
\textbf{Finding 1.} The optimal inspection interval \textbf{varies} based on the agent's execution health.
Compromised executions require frequent inspection to arrest failures (+9.68\% resolve rate), whereas healthy executions benefit from sparse inspection, which achieves a high resolve rate ($\ge$88.28\%) while reducing token overhead by 33.6\%.
\end{tcolorbox}


\subsection{Rule-based Guidance: Are Heuristics Sufficient for Misbehavior Detection?}
\label{sec:heuristic-sufficient}

\begin{wrapfigure}{r}{4.8cm}
  \centering
  \includegraphics[width=\linewidth]{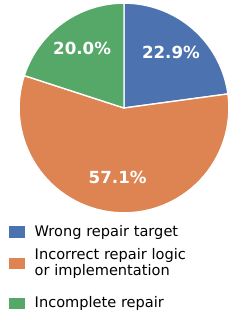}
  \caption{Categories of undetected errors.}
  \label{fig:undetected-errors}
\end{wrapfigure}

Existing rule-based monitors rely on fixed patterns to trigger interventions. To evaluate whether such heuristics are sufficient for coding agents, we equip mini-swe-agent with LivePlan, a state-of-the-art monitor that matches action sequences against predefined rules and generates LLM-guided advice when a rule fires.

On SWE-bench Lite, 152 out of 300 tasks failed. 
Notably, 35 of these failed tasks never triggered guidance delivery even once, representing critical silent failures where the monitor remains deactivated. To understand why these failures were overlooked, two experts conducted an open coding~\cite{opencoding} analysis of these trajectories, identifying three distinct categories of undetected misbehaviors (Figure~\ref{fig:undetected-errors}). A wrong repair target means the agent identifies an incorrect code region for modification. An incorrect repair logic or implementation indicates that although the agent locates the correct code region, it introduces faulty logic or buggy code. An incomplete repair signifies that the agent only partially addresses the issue, leaving remaining bugs unresolved~\cite{bug_really_fixed}. These misbehaviors primarily reflect internal reasoning flaws rather than surface-level execution anomalies~\cite{are-solved-issues-really-solved}, making them inherently difficult for predefined rules to detect without inspecting the agent's reasoning process.

\begin{tcolorbox}[colback=blue!5, colframe=blue!50!black, boxrule=0.5pt, left=6pt, right=6pt, top=4pt, bottom=4pt]
\textbf{Finding 2.} Monitors struggle to detect reasoning flaws when relying solely on predefined action patterns, causing $23.03\%$ of failed tasks ($35/152$) to remain completely unguided due to uncaptured diagnosis and repair errors.
\end{tcolorbox}

%% file: Chapters/Chapter3-Methodology.tex
\section{Methodology}

This section introduces our framework for real-time agent monitoring and strategic guidance. We begin by defining the core problem formulation, establishing our goal of tracking agent health to deliver timely interventions that keep agents aligned with their task objectives. We then present an overview of Cadence, our overall monitoring system, detailing its high-level architecture and its two principal components. Next, we delve into the two-tier intervention module, which determines the depth and nature of corrective guidance. Finally, we describe the inspection scheduler, which optimizes the frequency and timing of agent evaluations to balance monitoring precision with computational efficiency.

\subsection{Problem Formulation}

We first formally define the runtime monitoring task for coding agents. This problem comprises \textit{a base coding agent} tasked with resolving a software issue and \textit{an external monitor} supervising its execution.
The coding agent interacts with the development environment through sequential steps. At each step $i$, the agent reasons over the current context ($t_i$, thought), invokes tools ($a_i$, action), and receives execution feedback ($o_i$, observation)~\cite{react, codeact}. The agent's trajectory, which records $n$ steps, can be denoted as
\[
\tau = \langle (t_1, a_1, o_1), (t_2, a_2, o_2), \ldots, (t_n, a_n, o_n) \rangle.
\]
At run time, the monitor synchronously inspects the \textit{prefix trajectory} ($\tau_{\leq k}$) up to the current step $k$, where step $k$ presents the proposed thought $t_k$ and action $a_k$ without executing the $a_k$.
Upon each inspection, the monitor generates guidance $g_k$ based on the observed execution and integrates it as a distinct message into the trajectory.
For guidance appended after the current step, the resulting trajectory is represented as
\[
\tau' = \langle (t_1, a_1, o_1), \ldots, (t_k, a_k, o_k), g_k, (t_{k+1}, a_{k+1}, o_{k+1}), \ldots  \rangle,
\]
The coding agent then conditions on $\tau'$ in generating subsequent steps. The objective of runtime monitoring is to detect and mitigate misbehaviors, steering the agent toward successful task resolution.

\subsection{Cadence Overview}

Cadence consists of two modules: a two-tier intervention module and an inspection scheduler. The first module decides how to intervene, the second decides when to inspect, and together they provide strategic guidance. Specifically, \textit{the intervention module} evaluates recent execution health and applies an appropriate intervention level, either advisory or replacement, to the current step. For normal execution or only minor lapses, an advisory-level intervention allows the step to execute while appending guidance to the trajectory for future steps. For severe misbehaviors, a replacement-level intervention discards the proposed step before execution and prompts the agent to regenerate it with the guidance. \textit{The inspection scheduler} then uses this intervention level as an indicator of current execution health to adjust the inspection interval. This adaptive mechanism enables \textsc{Cadence} to back off during stable execution and intensify inspection during severe failures, preserving effectiveness while reducing monitoring overhead.

\begin{figure}[t]
    \centering
    \includegraphics[width=0.9\linewidth]{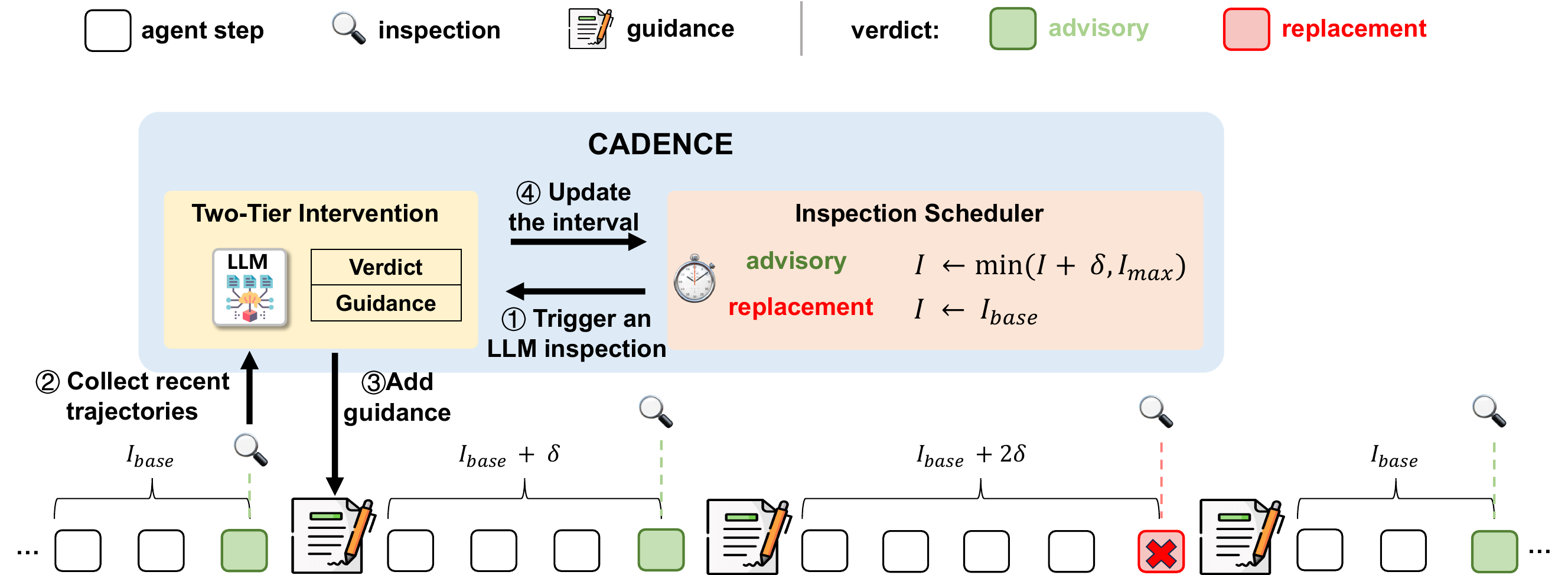}
    \caption{The overall workflow of Cadence.}
    \label{fig:cadence_overview}
\end{figure}

As illustrated in Figure~\ref{fig:cadence_overview}, the inspection scheduler initializes with a base inspection interval. Every time this threshold is reached, the scheduler triggers the intervention module to evaluate the trajectory since the last inspection. The intervention module then decides an intervention level and provides strategic guidance to the coding agent. The scheduling module then updates the inspection interval based on the level. This cycle continues until the coding agent completes the task. The following sections describe each module of Cadence in detail.

\input{Resources/Images/adaptive-interval-prompt}

\subsection{Two-Tier Intervention Module}
The execution health of a coding agent manifests across distinct severity levels during task execution. A minor issue, such as an occasional tool misuse, often does not lead to task failure, since the agent can self-correct in the subsequent step based on environment feedback. In contrast, a severe mistake, such as repeatedly failing the same attempt, can significantly damage the execution workflow and lead to task failure. Therefore, a uniform intervention policy struggles to handle both. 
An overly aggressive strategy introduces unnecessary disruptions, and a permissive strategy lacks the efficacy to break the coding agent out of critical, compounding failure modes.

To address this trade-off, Cadence adaptively determines how to intervene according to execution health. Instead of relying on rigid, hard-coded rules, Cadence leverages LLM-based semantic reasoning to evaluate execution health and dynamically select intervention level, applying advisory guidance for satisfactory health and replacement actions for compromised health.

At each inspection, the module first invokes LLMs to evaluate the trajectory segment since the last inspection point, assign an appropriate intervention level and generate corresponding guidance text. Figure~\ref{fig:adaptive-interval-prompt} illustrates the prompt used for this. Rather than using rigid and specific error patterns, we characterize execution health at a high conceptual level. Specifically, satisfactory execution health indicates that the coding agent is making steady progress toward resolving the issue, despite possible minor lapses in localized steps; compromised execution health signifies that the agent has drifted from the task goal, manifested by a persistent lack of progress or repeated failure modes. We deliberately avoid defining health through specific error types because different agents possess varying capabilities to resolve identical errors. Adopting high-level conceptual definitions thus enables Cadence to generalize across different coding agents.

Crucially, Cadence does not simply re-prompt a secondary LLM with the original failed problem. Instead, it uses task decomposition to reshape the problem representation~\cite{chain-of-thought, leasttomost, decomposed}. By decomposing a failing trajectory into localized execution health assessments, Cadence narrows the context window~\cite{lost-in-the-middle} and restricts the search space, which can steer the monitor toward more reliable regions of its representation manifold. This decomposition also shifts the prompt in distribution. Long, multi-step error recovery trajectories are rare and noisy in pre-training data, while explicit diagnostic reasoning, such as analyzing code execution logs and troubleshooting steps, is heavily represented in standard corpora like StackOverflow Q\&A threads and debugging forums. By converting an open-ended coding task into a structured execution health review, Cadence projects the problem onto a subspace where the monitor's pre-trained prior is stronger. This decomposition is complementary to monitor capability rather than a substitute for it. Different LLM backbones also contribute to monitor performance, as our results for Section~\ref{sec:rq2} show, because the decomposition changes what the monitor is asked to do, but it does not remove the need for a capable monitor.

\begin{figure}[t]
\centering
\includegraphics[width=0.9\linewidth]{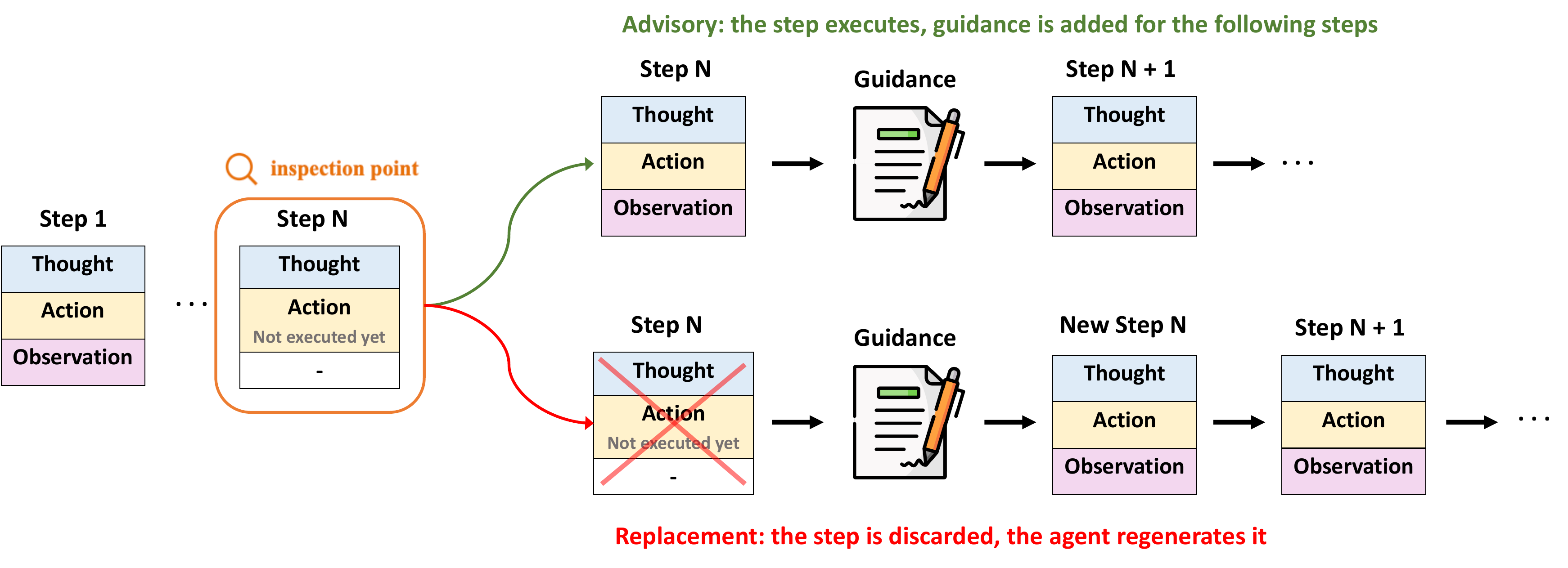}
\caption{Trajectory-level effect of the advisory and replacement branches.}
\label{fig:two-tier-mechanism}
\end{figure}

Figure~\ref{fig:two-tier-mechanism} illustrates the workflow of the two-tier intervention mechanism.  At each inspection point, it examines the recent trajectory segments, alongside the thought and action proposed for the current step, prior to execution. Based on the monitor's assessment, the module enforces one of two intervention policies.
Under an advisory intervention, the proposed action is permitted to execute, after which the generated guidance is appended to the trajectory as a user message to guide subsequent steps. Under a replacement intervention, Cadence discards the proposed step and injects the guidance into the context and prompts the agent to regenerate the step, preventing catastrophic failure modes from compounding. Finally, the chosen intervention level serves as an execution health signal, passed to the scheduling module.

\input{Resources/Algorithm/interval_update_algorithm}

\subsection{The Inspection Scheduler}

Cadence's scheduler dynamically changes the inspection interval based on the intervention module's feedback~\cite{congestion-control}, as formalized in Algorithm~\ref{alg:scheduling}. In the beginning, the module is configured with a base inspection interval $I_{base}$, a fixed growth step $\delta$, and a maximum inspection interval $I_{max}$ that prevents the interval from growing so large that inspections effectively stop. During execution, an inspection is triggered whenever the agent completes $I$ steps since the last check.
The scheduler then updates the interval according to the observed intervention level. 
When the intervention level is advisory, it indicates that the agent's execution health is stable and only requires light guidance. In this case, the module increases the inspection interval by the growth step, capped at the maximum interval (i.e., $I \leftarrow \min(I + \delta, I_{max})$). When the intervention level is replacement, however, it indicates that the agent is misbehaving seriously during execution. The module then resets the inspection interval to the base value (i.e., $I \leftarrow I_{base}$), providing tighter supervision to guide the agent back on track. Through this adaptive scheduling mechanism, the scheduling module maintains inspection effectiveness while limiting unnecessary monitoring burden.

%% file: Resources/Images/adaptive-interval-prompt.tex
\begin{figure}[t]
\centering
\begin{tcolorbox}[colback=gray!5, colframe=black!50, boxrule=0.5pt, fontupper=\small\ttfamily]
\raggedright
Input: \texttt{<issue\_description>}, \texttt{<recent\_guidance\_history>}, \texttt{<recent\_trajectory>}, \texttt{<proposed\_next\_action>} (the agent's proposed next action, not yet executed).

\textbf{[Decision Rules]}
Select exactly one verdict based on execution health:
\begin{itemize}[leftmargin=*, topsep=2pt, itemsep=2pt]
  \item \textbf{advisory}: Execution is on track or exhibits only minor deviations. The proposed action will be executed as planned. Provide forward-looking guidance applicable to subsequent steps.
  \item \textbf{replacement}: The agent is stuck, repeating a failed approach, or the proposed plan or action would make things worse. The proposed action will be discarded. Provide corrective guidance to regenerate this step.
\end{itemize}


Output:\\
\texttt{<verdict>advisory|replacement</verdict>}\\
\texttt{<guidance>Corrective instruction or strategic advice</guidance>}
\end{tcolorbox}
\caption{The prompt used by CADENCE's two-tier intervention module.}
\label{fig:adaptive-interval-prompt}
\end{figure}

%% file: Resources/Algorithm/interval_update_algorithm.tex
\begin{algorithm}[t]
\caption{Inspection Scheduling}
\label{alg:scheduling}
\begin{algorithmic}[1]
\Require base interval $I_{\text{base}}$, maximum interval $I_{\max}$, growth step $\delta$
\State $I \gets I_{\text{base}}$, $\text{last} \gets \textsc{None}$, $i \gets 0$ \Comment{$I$: current interval; last: index of the most recent inspection}
\While{task not finished}
    \State $i \gets i + 1$; agent proposes thought $t_i$ and action $a_i$
    \If{$(\text{last} = \textsc{None} \text{ and } i \geq I)$ \textbf{or} $(\text{last} \neq \textsc{None} \text{ and } i - \text{last} \geq I)$}
        \State $v \gets \textsc{TwoTierIntervention}(\tau, t_i, a_i)$
        \If{$v = \textit{advisory}$}
            \State execute $a_i$; inject guidance; $I \gets \min(I_{\max}, I + \delta)$
        \Else \Comment{$v = \textit{replacement}$}
            \State discard $(t_i, a_i)$; inject guidance; agent regenerates the step; $I \gets I_{\text{base}}$
        \EndIf
        \State $\text{last} \gets i$
    \Else
        \State execute $a_i$
    \EndIf
\EndWhile
\end{algorithmic}
\end{algorithm}

%% file: Chapters/Chapter4-Experimental_Setup.tex
\section{Experimental Setup}

We evaluate Cadence by answering the following research questions (RQs):
\begin{itemize}[leftmargin=*, itemsep=6pt, topsep=6pt, partopsep=0pt]
    \item \textbf{RQ1:} How effective and efficient is Cadence compared with existing runtime monitors?
    \item \textbf{RQ2:} How robust is Cadence across different backbone models and scheduling parameters?
    \item \textbf{RQ3:} How does each module contribute to the performance of Cadence?
    \item \textbf{RQ4:} When does Cadence succeed, and when does it fall short?
\end{itemize}

\subsection{Datasets}
\label{sec:datasets}
We conduct our experiments on SWE-bench Lite~\cite{swebench}, a benchmark of 300 issues collected
from popular open-source Python repositories on GitHub and widely used to evaluate
coding agents. Each issue provides a task description, a base commit, and a set of
tests. These tests fall into two groups. FAIL\_TO\_PASS tests fail on the base commit and should pass once the issue is resolved. PASS\_TO\_PASS tests already pass on the base commit and should remain passing afterwards. Given the task description, the coding agent runs on the repository at the base commit and produces a patch. An issue is resolved only if all tests in both groups pass after applying the patch, indicating it fixes the issue without introducing regressions~\cite{cure-worse-than-disease, GenProgAG}.

\subsection{Evaluation Subjects}

\subsubsection{Monitoring Baselines}
We compare Cadence against three monitoring baselines covering both periodic guidance and rule-based guidance described in Section~\ref{sec:introduction}.

\textbf{Fixed-Interval ($k$)} represents periodic guidance. Every fixed $k$ steps, it pauses the coding agent, invokes an LLM to inspect the recent trajectory, and appends the resulting guidance into the coding agent's context. This design is a synchronous version of Wink~\cite{wink}. Original Wink~\cite{wink} allows the coding agent to continue executing while it generates guidance in the background. Once finished, Wink appends the guidance to the current context. Because guidance generation speed and coding agent execution speed vary with LLM delays, this asynchronous design causes the guidance to be injected at unpredictable execution steps. We adopt a synchronous design to eliminate LLM latency as a confounding variable, ensuring a fair evaluation of intervention timing.

\textbf{LivePlan}~\cite{liveplan} represents rule-based guidance. It maps coding agent actions into predefined workflow phases and tracks state transitions using an execution graph. When predefined rules detect state anomalies, such as phase stalling or action loops, an LLM is prompted to generate targeted guidance. Depending on anomaly severity, LivePlan either blocks the erroneous step to force a regeneration with guidance, or simply appends guidance to context for subsequent steps.

\textbf{Static Rule-Based} represents another rule-based guidance. It uses the same heuristics as LivePlan but does not call an LLM. Instead, each rule has a fixed guidance message written in advance, which the monitor returns whenever that rule fires. This baseline shows how well the coding agent performs with general guidance that is not tailored to its current situation.

Additionally, we include \textbf{Vanilla}, the standalone agent without monitoring, as a control baseline to quantify the net benefits brought by monitoring.

\subsubsection{Working Agents}
To examine whether monitors generalize across coding agent scaffolds, we evaluate the monitors on two representative coding agents that manage actions differently. 

\textit{mini-swe-agent}~\cite{swe-agent} exposes only the bash command to the model. Every operation, such as reading files and editing code, is written as a shell command. Each command runs in a new subshell, so changes to the working directory or to environment variables do not carry over to later steps. The agent submits its solution by producing a git diff of the files.

\textit{Moatless}~\cite{moatless} instead exposes a set of 11 structured tools that cover search, file
reading, editing, test execution, and task completion. Before the run, Moatless
builds a vector index of the repository with a code embedding model. The agent can
then call SemanticSearch to locate relevant code with a natural language query,
instead of relying on exact text matching. In the end, Moatless invokes either VerifiedFinish to submit the patch, or Reject to report that the issue cannot be resolved.

\subsection{Evaluation Metrics}
We evaluate Cadence and the baselines along two dimensions, effectiveness and efficiency.

To evaluate \textbf{effectiveness}, we report \textit{Resolve Rate}, which is the standard metric used by SWE-bench Lite. It is defined as the percentage of tasks resolved, elaborated in Section~\ref{sec:datasets}.

To measure \textbf{efficiency}, we report two key metrics. \textit{Tokens per Task} denotes the average token count per task consumed by both the working agent and the monitor. \textit{Token Leverage} captures overall token-efficiency, defined as the number of resolved tasks per million total tokens consumed.

Additionally, for monitors using LLMs, we report \textit{Invocations per Task} as the average number of monitor LLM calls per task.

\subsection{Environment and Implementation}
We conduct all experiments in isolated Modal sandboxes based on SWE-bench's official prebuilt task images, following the default step limits established in prior studies (150 for mini-swe-agent and 50 for Moatless)~\cite{moatless, swe-agent}. Unless otherwise specified, Cadence uses a base interval of 5, a growth step of 2, and a maximum interval of 20 (i.e., reference setting).

To run the baselines on both agents, we make two adaptations for Moatless. The existing heuristic rules target mini-swe-agent only, so we construct a corresponding set of rules for Moatless that follow the same scope. These new rules are used by both Static Rule-Based and LivePlan. We also configure Moatless to run in linear exploration mode, which prevents the agent from reverting to a previous state to explore alternative actions and keeps each trajectory linear, consistent with the trajectory representation used throughout this paper.

For reproducibility, we set the temperature of every model to zero whenever the model supports it, which reduces randomness~\cite{llm-rondomness}. We also make all baseline implementations used in our experiments available in our repository.

%% file: Chapters/Chapter5-Evaluation.tex
\section{Evaluation}

We empirically evaluate Cadence across four research questions: overall effectiveness and efficiency (RQ1), robustness across configurations (RQ2), module contributions via ablation (RQ3), and trajectory analysis of successes and failures (RQ4).

\input{Resources/Experiment_Data/rq1_CADENCE_effectiveness}

\subsection{RQ1: How effective and efficient is Cadence compared with existing runtime monitors?}

In this section, we conduct a comprehensive evaluation of Cadence's effectiveness and efficiency against all baselines, using Gemini-2.5-Flash as the coding agent backbone model and GPT-5.2 as the monitor backbone model. Since GPT-5.2 is a reasoning model and does not support a temperature setting, we instead leave its reasoning effort at the default medium level. Regarding baseline configurations, prior studies typically adopt a fixed interval of 5 steps for the Fixed-Interval baseline. To ensure a thorough comparison and avoid biasing the baseline with a single interval choice, we also evaluate intervals of 3 and 7 steps.

\begin{figure}[t]
\centering
\includegraphics[width=0.8\linewidth]{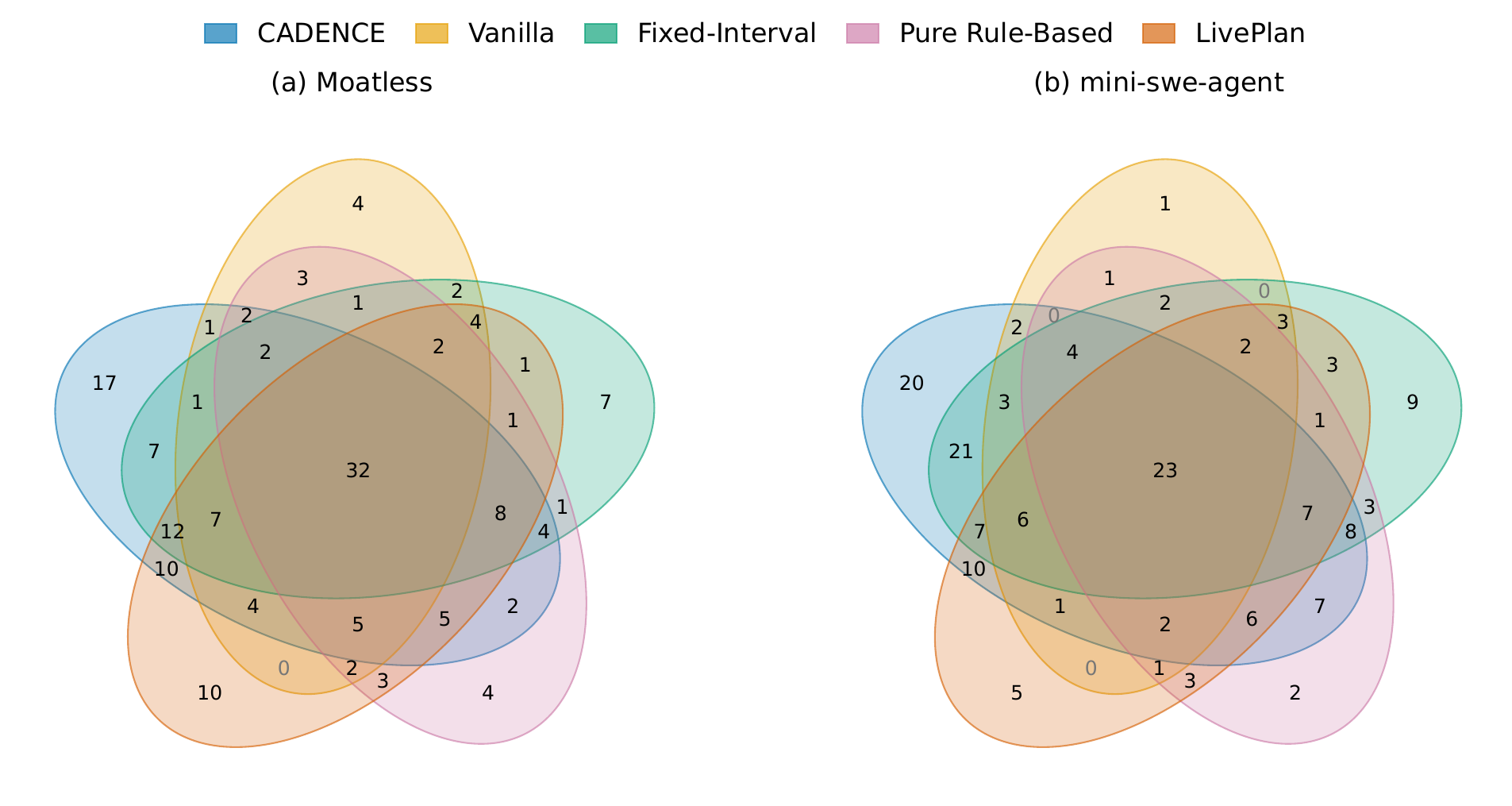}
\caption{Overlap in resolved tasks among Cadence and the four baselines. For visual clarity and focused comparison, the Fixed-Interval baseline is represented using $k=5$ as its primary configuration.}
\label{fig:rq1-venn}
\end{figure}

Table~\ref{tab:rq1-main} reports the main results. Cadence achieves the highest resolve rate across both coding agents. Specifically, it outperforms the strongest baseline on each agent, improving task resolution over LivePlan by 4.33\% (+13 tasks) on Moatless and over Fixed-Interval ($k=5$) by 8.33\% (+25 tasks) on mini-swe-agent. Compared with the Vanilla baseline, Cadence achieves absolute gains of 15.67\% on Moatless and 25.33\% on mini-swe-agent. As shown in Figure~\ref{fig:rq1-venn}, Cadence solves the most unique tasks on both agents, with 17 on Moatless and 20 on mini-swe-agent. Qualitative analysis shows that these uniquely resolved tasks are primarily rescued from incomplete repairs and incorrect bug root cause locations. The resolution of these complex misbehaviors benefits from Cadence leveraging LLM reasoning to track the agent's intent and evaluate its code modifications at appropriate frequency.

To further verify that these improvements are not attributable to chance, we conduct two-sided exact McNemar tests on the paired task-level resolution states, with Holm\cite{holm} correction across all baseline comparisons. The result shows that Cadence significantly outperforms all baselines on both coding agents except LivePlan on Moatless, where the adjusted $p$-value is 0.1175. This insignificance can be partially attributed to the distinct action spaces of the two agents. mini-swe-agent relies on flexible bash commands, making it harder for LivePlan to map actions to predefined misbehavior patterns accurately, where Cadence outperforms LivePlan by a larger margin of 15.67\%. Conversely, Moatless uses 11 structured tools with clearly defined functions, enabling more reliable mapping for LivePlan and narrowing the margin to 4.33\%. Even in this constrained setting, Cadence still achieves a higher resolve rate while maintaining superior token leverage (0.83 versus 0.69).

Alongside its effectiveness, Cadence also achieves efficiency gains. On mini-swe-agent, Cadence consumes the fewest tokens per task and yields the highest token leverage. This advantage stems from two primary design choices. First, Cadence extends the inspection interval when execution is healthy, thereby reducing monitoring overhead and keeping LLM invocation frequency second only to LivePlan, which triggers the LLM solely when heuristic rules fire. Second, empowered by LLMs, Cadence identifies a broader spectrum of misbehaviors and applies fine-grained interventions tailored to error severity, rescuing tasks that other baselines miss or mismanage. Consequently, Cadence resolves significantly more tasks for equivalent token expenditures, maximizing its overall token leverage. A similar trend holds on Moatless, where Cadence also achieves the top token leverage while maintaining average token consumption per task comparable to the strongest baseline, LivePlan (0.48M versus 0.51M).

\begin{tcolorbox}[colback=blue!5, colframe=blue!50!black, boxrule=0.5pt, left=6pt, right=6pt, top=4pt, bottom=4pt]
\textbf{Answer to RQ1.} Cadence consistently outperforms all baselines in task resolution while maintaining superior token efficiency. Across both coding agents, Cadence achieves the highest resolve rate (up to 42.33\%) and delivers the best token leverage, outperforming the strongest baseline by up to 105\% (0.41 versus 0.20).
\end{tcolorbox}

\subsection{RQ2: How robust is Cadence across different backbone models and scheduling parameters?}
\label{sec:rq2}

In this section, we investigate whether Cadence's performance advantages depend on specific system configurations. Specifically, we evaluate two key factors: the backbone models of both the coding agent and the monitor, and the hyperparameters controlling the inspection schedule. Across all experiments, we vary one factor at a time while holding all other settings at their default RQ1 configurations.

\subsubsection{Robustness to Backbone Models}
To evaluate Cadence's robustness across different LLM backbones, we systematically vary the base model of either the coding agent or the monitor while holding the other fixed at the default RQ1 setup (i.e., Gemini-2.5-Flash for the agent and GPT-5.2 for the monitor). Specifically, we test two alternative models for each role, one at a time, evaluating Cadence-monitored mini-swe-agent on SWE-bench Lite.

As shown in Table~\ref{tab:robustness-backbones}, Cadence consistently outperforms Vanilla across all evaluated backbone configurations for both the coding agent and the monitor.
For different monitor backbones, both alternative models achieve substantial performance gains, improving the resolve rate over Vanilla by at least 16.33\%. In contrast, for different coding agent backbones, although Cadence maintains a lead over Vanilla across all configurations, the magnitude of improvement varies. When paired with Qwen3-Coder-480B-A35B, Cadence yields a 15.00\% increase in resolve rate, whereas the gain is only 4.67\% for GPT-5-mini. This difference can be partly attributed to execution trajectory length. GPT-5-mini completes tasks in substantially fewer steps (26.6 vs. 60.0 for Qwen), offering the monitor fewer opportunities for inspection and intervention. 

\input{Resources/Experiment_Data/rq2_backbone_robustness}

\input{Resources/Experiment_Data/rq2_hypeparameters_robustness}

\subsubsection{Robustness to Scheduling Parameters}

Cadence incorporates three configurable parameters: the base interval, the growth step, and the maximum interval. We investigate parameter sensitivity to evaluate whether Cadence maintains robust performance across diverse configurations. 

\textit{Settings.} To isolate the impact of each parameter, we vary one parameter at a time. Since existing periodic monitors commonly adopt a 5-step interval, we evaluate base intervals clustered around this standard, selecting values of 3 and 7. For the growth step, we evaluate more aggressive expansion rates of 4 and 6. We omit the maximum interval from this analysis because it serves primarily as a safety fallback to prevent unbounded interval expansion; in practice, with only 1.7\% of inspection steps ever reached this limit across all experiments. For the remaining experimental configurations, we employ mini-swe-agent as the coding agent and retain the default backbone models from RQ1, specifically using Gemini-2.5-Flash for the agent and GPT-5.2 for the monitor.

Table~\ref{tab:robustness-intervals} presents the detailed results. Although performance exhibits modest variation across configurations, Cadence's lowest resolve rate (34.00\%, under a base interval of 5 and a growth step of 6) still matches the strongest baseline on mini-swe-agent, Fixed-Interval ($k=5$), while consuming nearly half the tokens per task (0.96M versus 1.75M). This finding demonstrates that Cadence's performance advantage is relatively stable across configurations. Furthermore, increasing either the base interval or the growth step generally reduces invocation frequency, whereas token leverage shows no clear dependence on these parameters.

\begin{tcolorbox}[colback=blue!5, colframe=blue!50!black, boxrule=0.5pt, left=6pt, right=6pt, top=4pt, bottom=4pt]
\textbf{Answer to RQ2.} Cadence demonstrates strong robustness across all settings, improving resolve rates by up to 20.33\% across diverse backbone models and matching the strongest baseline with 45.1\% fewer tokens even under its least favorable parameter configuration.
\end{tcolorbox}

\subsection{RQ3: How does each module contribute to the performance of Cadence?}

To evaluate the contribution of each component in Cadence, we conduct an ablation study on SWE-bench Lite. This evaluation uses the same backbone configuration as RQ1, pairing mini-swe-agent using Gemini 2.5 Flash with Cadence powered by GPT-5.2.

\input{Resources/Experiment_Data/rq3_ablation}

\subsubsection{Ablation of the Inspection Scheduling Module}

To evaluate the contribution of the scheduling module, we pair fixed inspection intervals with the two-tier intervention module, evaluating variants with intervals of 3, 5 and 7 steps, same value used by Fixed-Interval in RQ1.

Table~\ref{tab:rq3-ablation} summarizes the scheduling ablation results. For $k = 7$, the agent uses nearly the same token budget (1.025 versus 1.028 million tokens per task) as full Cadence with a similar inspection frequency (5.07 versus 5.90 invocations per task), yet its resolve rate drops by 9.00\%. This indicates that strategic inspection timing is crucial when monitoring overhead is held constant. Furthermore, setting $k = 5$ provides more monitoring resources, increasing inspections per task by 36.10\% and tokens per task by 17.03\% compared with full Cadence, yet yields virtually no resolution gain (42.00\% versus 42.33\%). This demonstrates that additional monitoring resources may fail to improve performance if inspection points are not carefully selected. Further increasing monitoring density with $k = 3$ provides the most abundant resources among all fixed variants, more than doubling the inspection invocations of full Cadence (12.45 versus 5.90). This setting raises the resolve rate by 3.33\%. However, this minor improvement is not statistically significant according to McNemar's test ($p = 0.24$). Moreover, $k = 3$ consumes 15.00\% more tokens and lowers token leverage by 6.30\% compared to full Cadence. These findings show that while heavy monitoring can slightly raise the resolve rate, its token efficiency still lags behind full Cadence. This confirms the core value of our dynamic scheduling module, which aims to allocate monitoring resources more effectively toward the steps where the agent needs them most.

\subsubsection{Ablation of the Two-Tier Intervention Module}

To evaluate the contribution of the two-tier intervention module, we construct two ablation variants that enforce a fixed intervention strategy for all inspections, using either always advisory or always replacement level actions. The monitor's original tier recommendation is retained solely by the scheduling module to update the inspection interval accordingly.

As shown in Table~\ref{tab:rq3-ablation}, resolve rates drop for both the always-advisory and always-replacement variants (by 7.33\% and 4.33\%, respectively) compared to the full Cadence, demonstrating that a single intervention strategy cannot adequately handle all execution states. In addition, the always-advisory variant exhibits longer average task steps and consumes 12.2\% more tokens than the always-replacement variant. This increase occurs partly because advisory interventions only provide guidance without directly interrupting the coding agent. Consequently, when an action introduces flawed code or deletes critical files, the agent requires additional steps to remediate the mistake, whereas replacement interventions directly halt erroneous actions and trigger immediate regeneration in a single step.

\begin{tcolorbox}[colback=blue!5, colframe=blue!50!black, boxrule=0.5pt, left=6pt, right=6pt, top=4pt, bottom=4pt]
\textbf{Answer to RQ3.} Both modules are critical to Cadence's overall performance. When the scheduling module is replaced with a fixed-interval schedule under a similar token budget ($k = 7$), the resolve rate drops by 9.00\%. Similarly, when the two-tier intervention module is replaced with a single fixed intervention strategy, the resolve rate drops by up to 7.33\%.
\end{tcolorbox}

\subsection{RQ4: When does Cadence succeed, and when does it fall short?}

To establish the operational scope of Cadence, we analyze its execution trajectories from the RQ1 experiments across both mini-swe-agent and Moatless. We evaluate the resolved and failed cases independently through case level examination. For resolved cases requiring intervention, we categorize the specific misbehaviors that Cadence corrects. For failed cases, we investigate the root failure mechanisms that constrain current monitor performance.

\subsubsection{Resolved Cases}
Two experts independently evaluated all 246 resolved cases to identify instances containing critical monitor interventions, defined as cases where the coding agent exhibited a misbehavior that would likely lead to task failure without monitor correction. This labeling achieved a Cohen's kappa of 0.94~\cite{cohen-kappa, k-explanation}, leaving 112 critical cases after resolving discrepancies through discussion, while the remaining 134 cases succeeded with auxiliary guidance that merely reinforced correct agent execution. Crucially, 90 of these 112 cases (80.4\%) failed under the vanilla agent setup, corroborating the necessity of monitor intervention for task resolution.

To analyze the misbehaviors addressed by Cadence, the two experts further performed a joint deductive-inductive~\cite{thematic-synthesis, Fereday2006DemonstratingRU} coding process on these 112 cases. They first mapped misbehaviors to existing categories from Section~\ref{sec:heuristic-sufficient}, and applied open coding to instances outside those definitions, yielding one new category termed \textit{tool-use error}. This category represents cases where the agent intends a valid code modification but fails during improper tool invocations, such as an edit command that silently fails or introduces syntax errors into the target file.

\begin{wrapfigure}{r}{4.8cm}
\centering
\includegraphics[width=\linewidth]{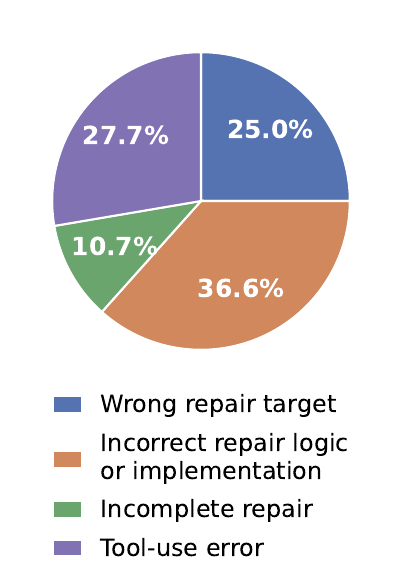}
\caption{Distribution of corrected misbehaviors across necessary monitor interventions ($n=112$).}
\label{fig:rq4-success-pie}
\end{wrapfigure}

As shown in Figure~\ref{fig:rq4-success-pie}, CADENCE effectively resolves both reasoning and tool use errors. Among these 112 corrected cases, replacement guidance serves as the decisive intervention in 88 cases (78.6\%), while advisory guidance resolves the remaining 24 cases (21.4\%), demonstrating the practical value of both intervention levels. Reasoning errors constitute the majority (72.3\%) of corrected misbehaviors, proving that detecting non-deterministic reasoning flaws is essential for trajectory monitoring. Tool use errors account for the remaining 27.7\%, occurring almost exclusively under mini-swe-agent (29 of 31 cases) because its free-form shell environment is inherently more error prone than Moatless's structured editing tools.

\begin{figure*}[t]
\centering
\includegraphics[width=\linewidth]{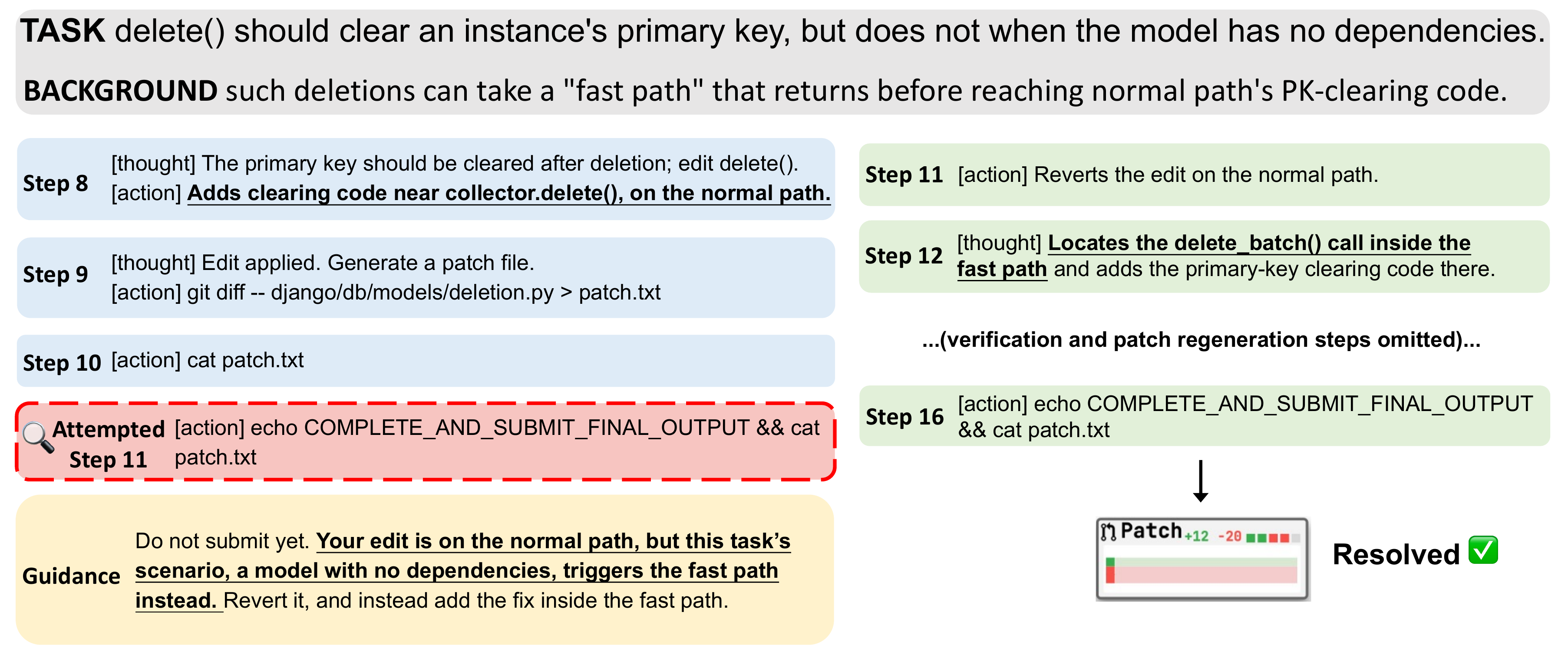}
\caption{A necessary Cadence intervention on django\_\_django-11179. Some observations and intermediate steps are omitted for brevity.}
\label{fig:rq4-case-study}
\end{figure*}

Figure~\ref{fig:rq4-case-study} illustrates how Cadence redirects a coding agent facing an incorrect repair target issue. In this task, the goal is to ensure that deleting a model instance without dependencies properly clears its primary key. Django implements deletion via two paths: a standard path and a fast path. During execution, the agent mistakenly added the clearing logic to the standard path, whereas models without dependencies actually trigger the fast path. Upon inspecting the agent's proposed submission, Cadence identified this mismatch and blocked the action, preventing a flawed patch from being submitted. Cadence then applied replacement guidance, discarding the submission and instructing the agent to implement the logic within the fast path instead. Guided by this intervention, the agent successfully modified the fast path and resolved the issue, whereas the baseline agent submitted an invalid patch that failed to satisfy the task requirements.

\subsubsection{Failed Cases}

We examine all failure cases from RQ1, excluding those caused by step budget exhaustion, leaving 284 cases for analysis. To systematically analyze these remaining failures, we first attribute each failure to either the coding agent or the monitor based on its primary root cause. A failure is attributed to the coding agent only when the monitor provides timely and correct guidance that the agent fails to carry out; all other cases are attributed to the monitor. Two authors with over two years of agent development experience independently perform this attribution, achieving a Cohen's kappa of 0.98. After mapping each case to its responsible party, we apply open coding in two stages to derive failure categories. First, the two authors jointly analyze a random sample of 100 cases, label each with a brief cause description, and iteratively merge similar labels until no new categories emerge. Second, the authors independently assign the remaining cases to these established categories, reaching a Cohen's kappa of 0.92. All disagreements in both stages are resolved through discussion.

Table~\ref{tab:rq4_failure-mechanisms} summarizes the distribution of these failure categories. Both the monitor and the coding agent contribute substantially to total failures (57.7\% versus 42.3\%). Among monitor failures, Root-Cause Misdiagnosis is the most prevalent category (26.8\%), encompassing instances where the monitor either identifies an incorrect defect location or focuses solely on a superficial cause. Inaccurate Guidance Generation (22.9\%) refers to cases where the monitor locates the root cause accurately, but delivers either a logic-flawed repair strategy or an overly abstract guidance plan. Inspection Timing Failure accounts for 8.1\% of cases, where short execution tasks finish before the first inspection triggers. This issue occurs exclusively on Moatless because it executes multiple tool calls in a single turn, which greatly compresses total trajectory steps. For coding agent failures, Incomplete Guidance Realization forms the largest overall category (39.4\%), representing cases where the agent accepts valid guidance but strays from the prescribed plan during multi-step execution. Finally, Guidance Non-Compliance is rare (2.8\%), describing scenarios where the coding agent remains entirely uninfluenced by the delivered guidance, showing no adjustment in its reasoning or actions.

These empirical findings point to three concrete directions for improving Cadence. First, to mitigate Inspection Timing Failure, the monitoring framework can incorporate tool invocation counts alongside step bounds when determining inspection timing for agents that bundle multiple actions per turn. Second, to address Root-Cause Misdiagnosis and Inaccurate Guidance Generation, prompt designs can be refined to encourage deeper root-cause reasoning; enforcing the monitor to thoroughly investigate underlying execution states rather than stopping at superficial symptoms can improve both diagnostic accuracy and the quality of generated guidance. Third, to address Incomplete Guidance Realization, the monitor can maintain an explicit checklist of raised issues to actively track resolution progress, gating final submissions until all flagged defects are fully addressed.

\input{Resources/Experiment_Data/rq4_failure}

\begin{tcolorbox}[colback=blue!5, colframe=blue!50!black, boxrule=0.5pt, left=6pt, right=6pt, top=4pt, bottom=4pt]
\noindent\textbf{Answer to RQ4:} Cadence succeeds by detecting and correcting both reasoning errors and tool use flaws. Conversely, Cadence falls short due to monitor limitations, such as untimely inspection, misdiagnosis, or inaccurate guidance, as well as the coding agent failing to completely execute the provided guidance.
\end{tcolorbox}

%% file: Resources/Experiment_Data/rq1_CADENCE_effectiveness.tex
\begin{table}[t]
\centering
\caption{Effectiveness and efficiency on SWE-bench Lite (300 tasks per agent). Parentheses show absolute resolution-rate gains over Vanilla in percentage points, and relative token-leverage changes over Vanilla in percent; token costs are reported in millions of tokens per task (M); a dash indicates no LLM-based monitor. \textbf{Bold} and \underline{underline} denote the best and second-best results, respectively.}
\label{tab:rq1-main}
\resizebox{\textwidth}{!}{%
\begin{tabular}{llccccc}
\toprule
Agent & Method & \makecell{Resolve\\Rate} & \makecell{Token\\Leverage} & \makecell{Tokens \\per Task } & Steps & \makecell{Invocations\\per Task} \\
\midrule
\multirow{7}{*}{mini-swe-agent} & Vanilla & 17.00\% & 0.12 & 1.42 & 53.2 & -- \\
 & Fixed-Interval ($k=3$) & 33.33\% (+16.33) & 0.14 (+17.0\%) & 2.38 & 68.9 & 20.8 \\
 & Fixed-Interval ($k=5$) & \underline{34.00\% (+17.00)} & 0.19 (+62.3\%) & 1.75 & 58.2 & 10.1 \\
 & Fixed-Interval ($k=7$) & 26.33\% (+9.33) & \underline{0.20 (+65.4\%)} & 1.33 & 50.8 & 6.0 \\
 & Static Rule-Based & 24.00\% (+7.00) & 0.18 (+46.3\%) & 1.37 & 57.8 & -- \\
 & LivePlan & 26.67\% (+9.67) & 0.18 (+51.5\%) & 1.47 & 58.5 & 2.2 \\
 & Cadence & \textbf{42.33\% (+25.33)} & \textbf{0.41 (+243.3\%)} & 1.03 & 46.4 & 5.9 \\
 \midrule
\multirow{7}{*}{Moatless} & Vanilla & 24.00\% & 0.75 & 0.32 & 24.8 & -- \\
 & Fixed-Interval ($k=3$) & 33.00\% (+9.00) & 0.69 (-8.3\%) & 0.48 & 29.2 & 9.1 \\
 & Fixed-Interval ($k=5$)& 30.67\% (+6.67) & \underline{0.79 (+4.8\%)} & 0.39 & 26.4 & 4.6 \\
 & Fixed-Interval ($k=7$) & 31.00\% (+7.00) & 0.74 (-1.6\%) & 0.42 & 28.1 & 3.5 \\
 & Static Rule-Based & 25.67\% (+1.67) & 0.56 (-25.6\%) & 0.46 & 28.1 & -- \\
 & LivePlan & \underline{35.33\% (+11.33)} & 0.69 (-7.6\%) & 0.51 & 28.9 & 2.2 \\
 & Cadence & \textbf{39.67\% (+15.67)} & \textbf{0.83 (+10.2\%)} & 0.48 & 29.3 & 4.0 \\
\bottomrule
\end{tabular}%
}
\end{table}

%% file: Resources/Experiment_Data/rq2_backbone_robustness.tex
\begin{table*}[t]
\centering
\small
\setlength{\tabcolsep}{5pt}
\caption{Cadence across different agent and monitor backbones, evaluated on 300 SWE-bench Lite tasks with mini-swe-agent. Rows without a monitor backbone represent Vanilla; all other rows use CADENCE. Parentheses show absolute resolution-rate gains over the matching Vanilla in percentage points, and relative token-leverage changes over the matching Vanilla in percent; tokens per task are reported in millions (M).}
\label{tab:robustness-backbones}
\resizebox{\textwidth}{!}{%
\begin{tabular}{lllccccc}
\toprule
& \makecell{Coding Agent\\backbone} & \makecell{Monitor\\backbone} & \makecell{Resolve\\Rate} & \makecell{Token\\Leverage} & \makecell{Tokens\\per Task} & Steps & \makecell{Invocations\\per Task} \\
\midrule
\multirow{2}{*}{Reference} & \multirow{2}{*}{Gemini 2.5 Flash} & -- & 17.00\% & 0.12 & 1.42 & 53.2 & -- \\
 & & GPT-5.2 & 42.33\% (+25.33) & 0.41 (+243.3\%) & 1.03 & 46.4 & 5.9 \\
\midrule
\multirow{4}{*}{Vary agent} & \multirow{2}{*}{\makecell[l]{Qwen3-Coder-\\480B-A35B}} & -- & 37.67\% & 0.45 & 0.83 & 56.5 & -- \\
 & & GPT-5.2 & 52.67\% (+15.00) & 0.55 (+20.9\%) & 0.96 & 60.0 & 7.2 \\
 \cmidrule(lr){2-8}
 & \multirow{2}{*}{GPT-5-mini} & -- & 48.33\% & 1.38 & 0.35 & 22.9 & -- \\
 & & GPT-5.2 & 53.00\% (+4.67) & 1.15 (-16.6\%) & 0.46 & 26.6 & 3.2 \\
\midrule
\multirow{2}{*}{Vary monitor} & \multirow{2}{*}{Gemini 2.5 Flash} & Claude Sonnet 4.6 & 37.33\% (+20.33) & 0.61 (+411.2\%) & 0.61 & 34.5 & 4.3 \\
 & & DeepSeek V4 Pro & 33.33\% (+16.33) & 0.35 (+193.1\%) & 0.95 & 42.7 & 5.6 \\
\bottomrule
\end{tabular}%
}
\end{table*}

%% file: Resources/Experiment_Data/rq2_hypeparameters_robustness.tex
\begin{table*}[t]
\centering
\small
\setlength{\tabcolsep}{4pt}
\caption{Sensitivity to the base interval and growth step on SWE-bench Lite tasks. The agent is mini-swe-agent with Gemini 2.5 Flash, and the monitor is Cadence with GPT-5.2; tokens per task are reported in millions (M). Bold parameter values denote the reference setting.}
\label{tab:robustness-intervals}
\resizebox{\textwidth}{!}{%
\begin{tabular}{ccccccc}
\toprule
Base interval & Growth step & Resolve Rate & Token Leverage & Tokens per Task & Steps & Invocations per Task \\
\midrule
\textbf{5} & \textbf{2} & 42.33\% & 0.41 & 1.03 & 46.4 & 5.9 \\
\midrule
5 & 4 & 37.00\% & 0.33 & 1.12 & 48.2 & 5.5 \\
5 & 6 & 34.00\% & 0.35 & 0.96 & 42.6 & 4.4 \\
\midrule
3 & 2 & 41.00\% & 0.31 & 1.32 & 48.8 & 9.1 \\
7 & 2 & 36.00\% & 0.32 & 1.12 & 47.0 & 4.6 \\
\bottomrule
\end{tabular}%
}
\end{table*}

%% file: Resources/Experiment_Data/rq3_ablation.tex
\begin{table}[t]
\centering
\caption{Ablation results on SWE-bench Lite (300 tasks). Parentheses show absolute resolution-rate changes over full Cadence in percentage points, and relative token-leverage and token-cost changes in percent; token costs are reported in millions of tokens per task (M). \textbf{Bold} denotes the best result.}
\label{tab:rq3-ablation}
\resizebox{\textwidth}{!}{%
\begin{tabular}{llccccc}
\toprule
Scheduling & Intervention & \makecell{Resolve\\Rate} & \makecell{Token\\Leverage} & \makecell{Tokens\\per Task} & Steps & \makecell{Invocations\\per Task} \\
\midrule
Dynamic & Two-tier & 42.33\% & \textbf{0.41} & 1.028 & 46.37 & 5.90 \\
\midrule
Fixed ($k=3$) & \multirow{3}{*}{Two-tier} & \textbf{45.67}\% (+3.33) & 0.39 (-6.3\%) & 1.182 (+15.00\%) & 47.95 & 12.45 \\
Fixed ($k=5$) & & 42.00\% (-0.33) & 0.35 (-15.2\%) & 1.203 (+17.03\%) & 50.72 & 8.03 \\
Fixed ($k=7$) & & 33.33\% (-9.00) & 0.33 (-21.0\%) & 1.025 (-0.30\%) & 45.35 & 5.07 \\
\midrule
\multirow{2}{*}{Dynamic} & Advisory only & 35.00\% (-7.33) & 0.31 (-23.7\%) & 1.114 (+8.33\%) & 48.09 & 6.80 \\
 & Replacement only & 38.00\% (-4.33) & 0.38 (-7.1\%) & 0.993 (-3.37\%) & 46.55 & 5.66 \\
\bottomrule
\end{tabular}%
}
\end{table}

%% file: Resources/Experiment_Data/rq4_failure.tex
\begin{table}[t]
\centering
\small
\caption{Primary failure attributions and categories across the 284 failure cases.}
\label{tab:rq4_failure-mechanisms}
\begin{tabular}{@{}llrr@{}}
\toprule
\textbf{Entity} & \textbf{Category} & \textbf{Count} & \textbf{\%} \\
\midrule
\multirow{3}{*}{Monitor}
& Inspection Timing Failure & 23 & 8.1 \\
& Root-Cause Misdiagnosis & 76 & 26.8 \\
& Inaccurate Guidance Generation & 65 & 22.9 \\
\midrule
\multirow{2}{*}{Coding Agent}
& Guidance Non-Compliance & 8 & 2.8 \\
& Incomplete Guidance Realization & 112 & 39.4 \\
\bottomrule
\end{tabular}
\end{table}

%% file: Chapters/Chapter6-Threats_to_Validity.tex
\section{Threats to Validity}

\textbf{Internal Threat.} Internal threats to validity mainly involve three aspects. First, to mitigate the randomness inherent in LLMs, we set the generation temperature to 0 whenever supported while leaving other settings at their defaults, and perform statistical significance testing on the main results. Second, to guarantee fair baseline comparisons, we utilize the official public implementations when available or re-implement the methods strictly following their original papers. Third, manual annotations during qualitative categorizations may introduce human subjectivity. To mitigate this threat, all trajectory misbehaviors and failure categories are determined by two annotators, with all discrepancies resolved through discussion.

\noindent\textbf{External Threat.} External threats to validity mainly involve two aspects. First, our evaluation focuses on bug fixing and feature implementation tasks in Python, whereas coding agents can also handle other kinds of tasks, such as test generation, and tasks in other programming languages. In the future, we plan to evaluate CADENCE on additional benchmarks, such as SWT-bench~\cite{swt-bench} and SWE-bench Pro~\cite{swe-bench-pro}, to further assess its generalizability. Second, there is a potential threat regarding whether the underlying LLMs were exposed to the evaluation tasks during training. Since all evaluated monitors share the same LLM backbones, any data contamination affects them comparably, ensuring that the relative performance gaps remain valid.

%% file: Chapters/Chapter7-RelatedWork.tex
\section{Related Work}
\label{sec:related_work}

Existing methods for enhancing coding agents primarily fall into two categories: offline learning and online monitoring. In the following, we review related works under both categories.

\subsection{Offline Learning}
Offline learning relies on historical execution records to enhance coding agents, primarily through model post-training or prompt-based experience reuse. Wei et al.~\cite{swe-rl} pioneered the use of reinforcement learning in coding agents by introducing SWE-RL, which fine-tunes the backbone model using rewards based on similarity to ground-truth patches. In addition, SWE-GYM~\cite{swe-gym} leverages powerful LLMs to perform coding tasks and uses the generated trajectories to train smaller models via imitation learning. For experience reuse in prompts, Wang et al.~\cite{Agent-Workflow-Memory} proposed Agent Workflow Memory, which extracts common workflows from past trajectories and integrates them into system prompts. Chen et al.~\cite{swe-exp} introduced SWE-Exp, which distills experiences from both successful and failed trajectories into a knowledge base and retrieves those relevant to a new task. However, these offline methods cannot handle runtime misbehaviors interactively; online monitoring fills this gap by providing real-time guidance during execution.

\subsection{Online Monitoring}

Online monitoring inspects runtime execution trajectories and delivers guidance upon detecting misbehaviors. Existing approaches fall into two categories based on their guidance triggering mechanisms: rule-based guidance and periodic guidance.

\subsubsection{Rule-based Guidance}
\label{sec:heuristic-based inspection}
Rule-based guidance predefines targeted misbehaviors and models specific misbehavior patterns over the agent's action space. A representative approach is LivePlan~\cite{liveplan}, which maps agent actions to predefined workflow phases to detect skipped or stalled execution phases. Additionally, LivePlan tracks execution trajectories using an action-graph representation to detect action loops. Upon matching patterns for severe misbehaviors, such as action loops or unauthorized phase skipping, LivePlan blocks the proposed action and invokes an LLM to generate guidance, prompting the agent to produce a new action. Conversely, for minor misbehaviors, it merely appends guidance to the trajectory upon the current step completion.

Through heuristic-based inspection, LivePlan achieves fast and precise misbehavior detection via pattern matching. However, heuristic rules fail to detect complex reasoning errors that require natural language understanding. Moreover, because each agent possesses a distinct action space, LivePlan requires manual adjustments to its action-phase mappings when deployed on a new agent.

\subsubsection{Periodic Guidance}
To detect broader categories of misbehaviors without requiring manual adaptation across different agents, recent approaches resort to LLM-based inspection. By exploiting the natural language understanding and domain knowledge of LLMs, such inspection can identify complex reasoning errors beyond the reach of heuristic rules. Furthermore, directly analyzing raw trajectory text independent of the agent's action space provides better generalizability across diverse coding agents. To keep cost under control, such monitors typically invoke LLM-based inspection periodically.

A representative approach is Wink~\cite{wink}. It invokes an LLM to inspect the execution trajectory at fixed step intervals. To mitigate the latency associated with frequent LLM calls, it performs inspections asynchronously, allowing the agent to continue execution without waiting for inspection results. Guidance is injected into the trajectory only when misbehavior is detected. Although this asynchronous design reduces runtime delays, a severe misbehavior may rapidly cascade into an unrecoverable failure while the agent continues executing before the inspection completes. More fundamentally, the inspection interval in Wink is statically predetermined and remains indifferent to the agent's real-time execution status. Consequently, a healthy agent and a misbehaving agent receive the same frequency of supervision, which represents a sub-optimal design choice as demonstrated by our motivating study (Section~\ref{sec:motivating study}).

%% file: Chapters/Chapter8-Conclusion.tex
\section{Conclusion}
In this paper, we propose Cadence, a dynamic runtime monitor for coding agents that integrates a two-tier intervention module with an inspection scheduler. The intervention module provides advisory guidance when agent execution health remains satisfactory, while issuing replacement guidance when severe misbehaviors occur. Based on the resulting intervention level, the scheduler dynamically adjusts the inspection frequency, increasing supervision for replacement-level cases and reducing it for advisory-level ones. Evaluation on 300 SWE-bench Lite tasks across two coding agents shows that Cadence achieves the highest resolve rate among all evaluated monitors while maintaining competitive token efficiency.

\section*{Data Availability}
All code for Cadence and baseline monitors, as well as the experimental results for each setup, are publicly available at \url{https://github.com/LogAnalysisTech/Cadence}.